\documentstyle[12pt,epsf]{article}

\begin{document}
\begin{flushright}
IFUP--TH 51/97 \\
UCY--PHY--97/08 \\
IFUM 587/FT
\end{flushright}
\vskip 2cm

\centerline{{\bf A critical comparison of different definitions 
of topological charge on the lattice}\footnote {Partially 
supported by EC Contract CHEX-CT92-0051 and by MURST.}}
\vskip 5mm
\centerline{B. All\'es$^{\rm a}$, M. D'Elia$^{\rm b,c}$, 
A. Di Giacomo$^{\rm b}$, R. Kirchner$^{\rm b,d}$}
\vskip 5mm
\centerline{$^{\rm a}${\it Dipartimento di Fisica, Sezione Teorica, 
Universit\`a di Milano and INFN,}} 
\centerline{\it Via Celoria 16, 20133--Milano, Italy}
\vskip 3mm
\centerline{$^{\rm b}${\it Dipartimento di Fisica dell'Universit\`a and INFN, 
Piazza Torricelli 2, 56126--Pisa, Italy}}
\vskip 3mm
\centerline{$^{\rm c}${\it Department of Natural Sciences, 
University of Cyprus, }}
\centerline{\it P.O. Box 537, Nicosia CY--1678, Cyprus}
\vskip 3mm
\centerline{$^{\rm d}${\it Institut f\"ur theoretische Kernphysik, 
Nussallee 14--16, 53115 Bonn, Germany}}

\vskip 2cm

\begin{abstract}
A detailed comparison is made between the field--theoretic and geometric
definitions of topological charge density on the lattice. Their 
renormalizations with respect to continuum are analysed. The definition
of the topological susceptibility $\chi$, as used in chiral Ward 
identities, is reviewed. After performing the subtractions required by it,
the different lattice methods yield results in agreement with each other.
The methods based on cooling and on counting fermionic zero modes
are also discussed.
\end{abstract}

\vfill\eject

\section{Introduction}

 The definition of topological charge density and of topological
susceptibility on the lattice has by now a long story with contrasting
results~\cite{cite1,cite2}. This paper intends to be a contribution
to clarify the issue. 

Lattice is a regulator of the theory. It should reproduce continuum
physics in the limit in which the cutoff is removed, i.e., in the limit
in which the lattice spacing $a$ tends to zero. Like any other regularization
scheme, however, appropriate renormalizations have to be performed to
determine physical quantities. Within the rules of renormalization
theory the topological charge density and its correlation functions
can be defined on the lattice with the same rigour as for any other
operator of the theory.

In QCD 
\begin{equation}
 Q(x) = {{g^2} \over {64 \pi^2}} \epsilon_{\mu \nu \rho \sigma} 
 F^a_{\mu \nu} (x) F^a_{\rho \sigma} (x).
\label{eq:qdef}
\end{equation}
$Q(x)$ has a fundamental physical role, being the anomaly of the 
$U_A(1)$ singlet axial current
\begin{eqnarray}
 \partial_\mu J^5_\mu(x) &=& -  2 N_f Q(x), \nonumber \\
 J^5_\mu(x) &=& \sum_{i=1}^{N_f} \overline\psi_i(x) 
\gamma_\mu \gamma_5 \psi_i(x).
\label{eq:anomaly}
\end{eqnarray}
$N_f$ is the number of light flavours. Eq.(\ref{eq:anomaly}) provides
a solution to the $U_A(1)$ problem of Gell--mann's quark model in which
$J^5_\mu$ is conserved and the corresponding $U_A(1)$ is a symmetry,
whereas in hadron physics neither parity doublets are observed, which
would correspond to a Wigner realization, nor the inequality $m_{\eta'}
\leq \sqrt{3} m_\pi$ is satisfied, which would correspond to a
spontaneous breaking \`a la Goldstone.

 Eq.(\ref{eq:anomaly}) could explain the higher value of $m_{\eta'}$
as suggested by an approach based on $1/N_c$ expansion of the theory. At the
leading order the anomaly is absent being $O(1/N_c)$, and $U_A(1)$
is a Goldstone symmetry like axial $SU_A(3)$. The idea behind this 
expansion is that already at this order the theory describes the main
physical features of hadrons (e.g. confinement)~\cite{cite3}. In the $1/N_c$
expansion the anomaly acts as a perturbation, displacing the pole
of the $U_A(1)$ Goldstone boson to the actual mass of the $\eta'$.
The prediction is~\cite{cite4,cite5}
\begin{equation}
 \chi = {{f_\pi^2} \over {2 N_f}} 
 \left( m_\eta^2 + m_{\eta'}^2 - 2 m_K^2 \right)
\label{eq:mass}
\end{equation}
where
\begin{equation}
 \chi = \int \hbox{d}^4x \langle 0 | T(Q(x) Q(0)) | 0 \rangle
\label{eq:chi}
\end{equation}
is the topological susceptibility of the vacuum in the unperturbed
($N_c = \infty$) theory. This means, among other facts, quenched
approximation, fermion loops being $O(g^2 N_f) \sim O(N_f/N_c)$.

 In fact, as we shall discuss in detail below, $\chi$ in 
Eq.(\ref{eq:chi}) is not defined if the prescription is not specified
for the singularity
of the product $Q(x)Q(0)$ as $x \longrightarrow 0$.
In refs.~\cite{cite4,cite5} the prescription 
which leads to Eq.(\ref{eq:mass}) is the following
\begin{equation}
 \chi = \int \hbox{d}^4(x-y) \partial^x_\mu \partial^y_\nu 
\langle 0 | T(K_\mu(x) K_\nu(y)) | 0 \rangle
\label{eq:prescription}
\end{equation}
where $K_\mu(x)$ is the Chern current 
\begin{equation}
 K_\mu = {g^2 \over {16 \pi^2}} \epsilon_{\mu\nu\rho\sigma}
 A^a_\nu \left(\partial_\rho A^a_\sigma - {1 \over 3}
 g f^{abc} A^b_\rho A^c_\sigma \right)
\label{eq:k}
\end{equation}
related to $Q(x)$ by the equation
\begin{equation}
 \partial_\mu K_\mu(x) = Q(x)
\label{eq:gradient}
\end{equation}
The prescription Eq.(\ref{eq:prescription}) eliminates all the 
$\delta$--like singularities in the product $K_\mu(x) K_\nu(y)$
as $x \longrightarrow y$.
In any regularization scheme the prescription Eq.(\ref{eq:prescription})
only leaves a multiplicative renormalization $Z^2$ for $\chi$, $Z$ being
the possible renormalization of $Q(x)$. 
Eq.~(\ref{eq:gradient}) implies that the total topological charge
\begin{equation}
 Q \equiv \int \hbox{d}^4x Q(x)
\label{eq:totalQ}
\end{equation}
has integer values. 

The regularized version of $Q(x)$, $Q_L(x)$, does not obey in general
(\ref{eq:gradient})~\cite{cite6}. According to the general rules
of renormalization theory (in pure gauge theory) 
\begin{equation}
 Q_L(x) = Z Q(x) .
\label{eq:QLZQ}
\end{equation}
In general $Z\not= 1$ unless Eq.~(\ref{eq:gradient}) is preserved 
by regularization. To determine $Z$ it is sufficient to measure
$\langle Q_L \rangle = a^4 Z \langle Q \rangle $ on a state belonging
to a definite eigenvalue of $Q$ (see section 2).

The product $Q_L(x) Q_L(y)$ will not satisfy the prescription
Eq.~(\ref{eq:prescription}) in general at the singularity 
$x \longrightarrow y$. In the limit $a \longrightarrow 0$ it will differ
from it by additive terms, $\delta\chi$, 
which can be classified by use of the
Wilson operator product expansion~\cite{cite7}. Defining ($V$ is the
4--volume)
\begin{equation}
 \chi_L \equiv {1 \over V} \sum_{x\;y} Q_L(x) Q_L(y)
\label{eq:chiLdef}
\end{equation}
we will have 
\begin{equation}
 \chi_L = {1 \over V} Z^2 Q^2 a^4 + \delta\chi
\label{eq:deltachi}
\end{equation}
where the first term corresponds to the prescription 
Eq.~(\ref{eq:prescription}). Taking the v.e.v. of Eq.~(\ref{eq:deltachi}) gives
\begin{equation}
 \chi_L = a^4 Z^2 \chi + \chi_0
\label{eq:reg1}
\end{equation}
with 
\begin{equation}
 \chi_0 = \langle 0 | \delta\chi | 0 \rangle . 
\label{eq:reg1primo}
\end{equation}
Taking the expectation value of Eq.~(\ref{eq:deltachi}) on eigenstates
$|q_n \rangle$ of $Q$ gives
\begin{equation}
 \langle q_n | \chi_L | q_n \rangle = {1 \over V} Z^2 q_n^2 a^4 +
 \langle q_n | \delta\chi | q_n \rangle .
\label{eq:qn}
\end{equation}
It is a generally accepted wisdom that renormalization effects produced
by short range quantum fluctuations are practically independent of 
the semiclassical instanton background which determines $q_n$. The 
independence on $q_n$ of $\langle q_n | \delta\chi | q_n \rangle$
can be checked numerically by Eq.~(\ref{eq:qn}) and proves to be
true within errors~\cite{cite13}. Then 
$\langle q_n | \delta\chi | q_n \rangle = \chi_0$, 
and $\chi_0$ can be determined from
Eq.~(\ref{eq:qn}) as $\langle q_n = 0 | \chi_L | q_n = 0 \rangle$, i.e. as
expectation value of $\chi_L$ on the trivial topological sector.

 From Eq.(\ref{eq:reg1})
\begin{equation}
 \chi = {{\chi_{\rm reg} - \chi_0} \over {Z^2 }}.
\label{eq:reg2}
\end{equation}
It is with this prescription that $\chi$ is expected to be 
\begin{equation}
 \chi = (180 \;\hbox{MeV})^4 
\label{eq:180}
\end{equation}
in the quenched approximation within an $O(1/N_c)$ systematic error.

 In this paper we will show that, if the prescription 
Eq.(\ref{eq:prescription}) is properly implemented, all methods
which have been proposed to determine $\chi$ on the lattice
give the same result. We shall do this by comparing the geometric
method~\cite{cite8,cite9} to define $Q(x)$ 
to the field--theoretic one~\cite{cite6,cite7} for $SU(2)$
gauge theory.
The same procedure, applied to $SU(3)$ indeed 
confirms~\cite{cite10} the expectation Eq.(\ref{eq:180}).

\section{Defining $Q(x)$ on the lattice}

 In analogy to any lattice operator, $Q_L(x)$ will be defined by the
requirement that, in the formal (na\"{\i}ve) limit 
$a \longrightarrow 0$ 
\begin{equation}
 Q_L(x) {\buildrel {a \rightarrow 0} \over \sim} a^4 Q(x) + O(a^6).
\label{eq:naive}
\end{equation}
A prototype definition is
\begin{equation}
Q_L(x) = {{-1} \over {2^9 \pi^2}} 
\sum_{\mu\nu\rho\sigma = \pm 1}^{\pm 4} 
{\tilde{\epsilon}}_{\mu\nu\rho\sigma} \hbox{Tr} \left( 
\Pi_{\mu\nu}(x) \Pi_{\rho\sigma}(x) \right).
\label{eq:qlattice}
\end{equation}
${\tilde{\epsilon}}_{\mu\nu\rho\sigma}$ is the
standard Levi--Civita tensor for positive directions while for negative
ones the relation ${\tilde{\epsilon}}_{\mu\nu\rho\sigma} =
- {\tilde{\epsilon}}_{-\mu\nu\rho\sigma}$ holds. 

 $O(a^6)$ irrelevant terms in Eq.(\ref{eq:naive}) will dissapear
in the scaling regime. However their presence may be used to improve
the operator~\cite{cite11}. In what follows $Q_L^{(i)}$, $(i=0,1,2)$
will denote the operator defined by Eq.(\ref{eq:qlattice}) and the
once and twice improved versions of it respectively. Improvement
is the recursive smearing of the links developed in~\cite{cite11}.
Also the geometric definition $Q_L^{\rm geom}(x)$ 
satisfies Eq.(\ref{eq:naive})~\cite{cite8}.

 Like any other regularized operator, $Q_L(x)$ will mix 
in the continuum limit, when irrelevant terms become unimportant,
with all the operators having the same quantum numbers and lower
or equal dimension. The only pseudoscalar of dimension $\leq 4$ 
being $Q(x)$ itself,
\begin{equation}
 Q_L(x) = Z Q(x).
\label{eq:z}
\end{equation}
The na\"{\i}ve expectation for $Z$ would be $Z=1$ 
since $Q$, as an integer, should not renormalize. 
As first realised in~\cite{cite6} this is not true on the
lattice where $Q_L(x)$ is not a divergence.
$Z$ can be computed in perturbation theory, as it was done in the early
works on the subject~\cite{cite6}. 
A better way is to measure $\langle Q_L \rangle$,
$Q_L$ being the total topological charge on the lattice,
on a state on which $Q$ has a known value, e.g. on a one--instanton
state where $Q=1$. This can be done by a heating technique~\cite{cite13}
where a
background instanton is put by hand on the lattice, and quantum
fluctuations at a given value of $\beta=2N_c/g^2$ are added to it. 
In the continuum the instanton configuration is stable, being
a minimum of the action, and therefore perturbing it by small
fluctuations does not change the value of $Q$. On the lattice
instantons are not stable, so that $Q$ could change during this
heating procedure. The way to avoid this inconvenience is to create a sample
of configurations by the usual Monte Carlo updating, starting from
the original instanton. Each of them is checked in its instanton
content by a rapid cooling: configurations where the original topological
charge seems to be changed are discarded. This can be done after
any number of heating steps and the result must be independent
of this number. The result is a sample with topological charge $Q$.
$Q_L$ measured on this ensemble 
will reach a plateau in this heating procedure, on which 
$Z$ can be read. If the instanton were stable, the plateau would
stay flat forever.

 This procedure has been checked and used repeatedly within the
field--theoretical method~\cite{cite10,cite14,cite15,cite16}. 
The result for $Z$ is shown in Fig.~1
as a function of $\beta$ for different definitions of $Q_L(x)$.
The data for the 0--, 1-- and 2--smeared operators are taken
from ref.~\cite{cite16} and are computed on a 1--instanton 
configuration on a $16^4$ lattice. The data for the geometrical 
definition are new and are computed with the same procedure.
For the geometric definition $Z$ is compatible with 1
(within two standard deviations).
However
the values of $Q_L^{\rm geom}$ have a large spread (as large as
$Q\pm 10$) showing that
it can assume values different from the original $Q$ on configurations
which are presumed to belong to that sector. Only the average satisfies
$\langle Q_L \rangle=Q$, but not the value configuration by
configuration. Moreover, on a given configuration the value of
$Q_L^{\rm geom}$ depends on the interpolation used to define 
it~\cite{cite9}.

\section{The topological susceptibility}

 The lattice topological susceptibility is written as
\begin{equation}
 \chi_L = \sum_x \langle Q_L(x) Q_L(0) \rangle = {Q_L^2 \over V}
\label{eq:chilattice}
\end{equation}
and analogously for $Q_L^{\rm geom}$.
To make connection to the continuum susceptibility as defined by
Eq.(\ref{eq:prescription}), in general there will be an additive
renormalization due to the singularity at $x \longrightarrow y$ 
and a multiplicative residual renormalization $(x \not= y)$
which will simply be the square of $Z$ computed in the previous section.

As a matter of fact $\langle Q_L(x) Q_L(0) \rangle$ is expected to be
negative due to reflection positivity at $x\not=0$ since $Q_L(x)$
changes its sign under time reversal~\cite{cite17}. In fact this holds 
at distances larger than the extension of the operator if it is
smeared. Figs.~2~and~3 show that this is indeed the case both for the
geometric operator and the field--theoretical definition. Since
$Q_L^2$ is positive, its value is determined mainly by the point
at $x=0$, i.e. by the singularity of the product at 
$x \longrightarrow 0$. This peak is there, no matter how
$Q_L(x)$ is defined, and its height depends on the definition used.
In Figs.~2~and~3 the values for $\langle Q_L(x) Q_L(0) \rangle$
have been summed over all points $x$ inside 
a shell at distance $|x|$ from the origin $x=0$.
The width of this shell was $1.2\;a$.

Thus in general~\cite{cite6,cite7}
\begin{equation}
 \chi_L = Z(\beta)^2 a^4 \chi + M(\beta).
\label{eq:chil}
\end{equation}
$M(\beta)$ will describe a mixing with all scalar operators of
dimension $\leq 4$ ($\overline\beta(g)$ is the beta function),
\begin{equation}
 M(\beta) = A(\beta) \langle {{\overline\beta(g)} \over g}
                      F^a_{\mu\nu} F^a_{\mu\nu} \rangle a^4 +
            P(\beta) \cdot 1.
\label{eq:mixing}
\end{equation}
$M$ is the value of $\chi_0$ in Eq.(\ref{eq:reg1}) in the lattice
regularization.

To match the prescription of Eq.(\ref{eq:prescription}), $\chi$
has to be zero in the sector $Q=0$. In that sector thus 
$\chi_L = M(\beta)$ and $M(\beta)$ can be determined by measuring
$\chi_L$ in it. This is again done by a heating procedure~\cite{cite13}.
The flat, zero field configuration, ($U_\mu(x)=1$) can be dressed
with local quantum fluctuations, which do not change its topological 
content, by the usual updating procedure at the desired value of
$\beta$. $\chi_L$ will soon reach a plateau: if the sector were
stable the plateau would persist forever. Instead non--vanishing topological
charge can be created on the lattice and care must be
taken to eliminate configurations where this happens. Again this
must be done by cooling and checks can be done to test the 
consistency of the procedure. Fig.~4 shows the determination of
$M(\beta)$ for the geometric definition and for three different
field--theoretical definitions. Analogously to Eq.(\ref{eq:reg2}), 
from Eq.(\ref{eq:chil}) we obtain
\begin{equation} 
 a^4 \chi = {{\chi_L - M(\beta)} \over Z(\beta)^2}.
\label{eq:chif}
\end{equation}
$\chi_L$, $M$ and $Z$ depend on the choice of the regulator
as well as on the choice of the action; 
$a$ depends on the choice of the action
but $\chi$ must be independent of all of it.
 Fig.~5 shows that this is the case. 
In this figure we have used the data of ref.~\cite{cite16} for the
0-- and 2--smeared field--theoretical charges.
The data for the geometric definition has been obtained on a 
$16^4$ lattice with the same updating procedure (heat bath) and
compatible statistics (5000 configurations).
The result of the simulations is in fact $\chi/\Lambda_{L}^4$.
Usually people determine $\Lambda_{L}$ by computing 
the string tension $\sigma/\Lambda_{L}^2$ and by assuming 
the physical value for $\sigma$. This allows to express $\chi^{1/4}$
in physical units. We do same in order to compare our result
with other people's determinations.
The scale is determined from the data of~\cite{cite16bis,cite16tris}. 
The data at 2--smearings yield $(\chi)^{1/4}=198\pm 2\pm 6$ MeV
for $SU(2)$ gauge group,
the first error being statistical and the second one coming from the
error in $\Lambda_L$.
The ``na\"{\i}ve'' unsubtracted
geometric definition does not scale and is almost one order of magnitude
larger than the subtracted value. In the jargon of the geometrical
method this is called an effect of dislocations. 
It is the mixing to the identity operator which 
indeed describes these dislocations, having dimension lower than 4.
There is however 
an additional mixing in Eq.(\ref{eq:mixing}) which has
the same dimension as $\chi$ and still must be subtracted: checking
only by dimension is not sufficient to ensure that $\chi_L$ is
indeed equal to the physical $\chi$, as defined by 
Eq.(\ref{eq:prescription}). 

 Figs.~6~and~7 show the distribution of
values for $Q_L$ in the sector with trivial topology. 
Its variance is, apart from a normalization factor,
a measure of $M(\beta)$.
A good operator $Q_L(x)$ is one for 
which the subtraction $M(\beta)$ is small compared to $\chi_L$.
On the other hand, also having $Z \approx 1$ is more reassuring 
than having a small $Z$.

 Table I shows $\chi_L$, $Z$ and $M$ for the 0--, 1--, 2--smeared
field--theoretical charges and the geometric charge at $\beta=2.57$.

 The 2--smeared definition of $Q_L(x)$ is the best among these choices.
The geometric definition is good with respect to $Z$ but 
is definitively bad with respect to the additive renormalization $M$.

\section{Discussion}

 The main conclusion of the above analysis is that with any definition
of topological charge density on the lattice, an additive
renormalization for the topological susceptibility and a multiplicative
one are necessary. If properly 
renormalized all definitions bring about the same
physical value for $\chi$.

 Confusion on this subject in the past was generated by a mistreatment of 
renormalization. On the one hand, the geometric definition was
believed to be free from renormalizations because it gave
always integer values for the total topological charge. 
This seems to be true for the multiplicative
renormalization. Having integer values, however, does not exempt
from having singularities at short distance in the product which
defines $\chi_L$. Fig.~5 is clearly proving that.

 The field--theoretical definition started as a na\"{\i}ve definition.
$Z$ was not noticed and put equal to 1; $P(\beta)$ was subtracted by
use of perturbation theory. As a result $Z^2\chi$ was determined
instead of $\chi$ itself, and found to be much smaller than 
the expectation Eq.(\ref{eq:180})~\cite{cite18}.

 The idea was then put forward that the na\"{\i}ve
definition might not be correct and the geometric method~\cite{cite8,cite19},
the cooling method~\cite{cite20,cite21} and the Atiyah--Singer based 
methods~\cite{cite22} were developed.
The na\"{\i}ve method was promoted to field--theoretic method only after
introducing $Z$ and a correct subtraction $M$~\cite{cite6,cite7}. 
The non--perturbative
determinations of these constants~\cite{cite13} 
as explained above, finally brought
about a reliable determination of $\chi$, which is indeed regulator
independent.

 The cooling method authomatically performs the additive subtraction 
because it gives $\chi_L=0$ on the trivial sector; and also brings
$Z$ to 1 by freezing the quantum fluctuations. The problem with it
was that instantons could be lost in the procedure, leading to an
underestimation of $\chi$. Cooling with improved forms of the 
action~\cite{cite23,cite23bis}
seems to have eliminated this problem and gives indeed results which
confirm the field--theoretic determination.
The same seems to be true for the modern versions of the Atiyah--Singer
procedure~\cite{cite24}.

\section{Acknowledgements}

We are greatly indebted to CNUCE (Pisa) for qualified technical 
assistance in the use of their IBM--SP2 and to the CRT Computer Center of 
ENEL (Pisa) for warm hospitality and collaboration in the use of their 
Cray YMP--2E. We also thank Gerrit Schierholz for providing us with the
vectorized fortran code for the geometrical topological charge and
Pietro Menotti for clarifying discussions about the reflection
positivity.

\vskip 15mm


\newpage

\noindent{\bf Figure captions}

\begin{enumerate}

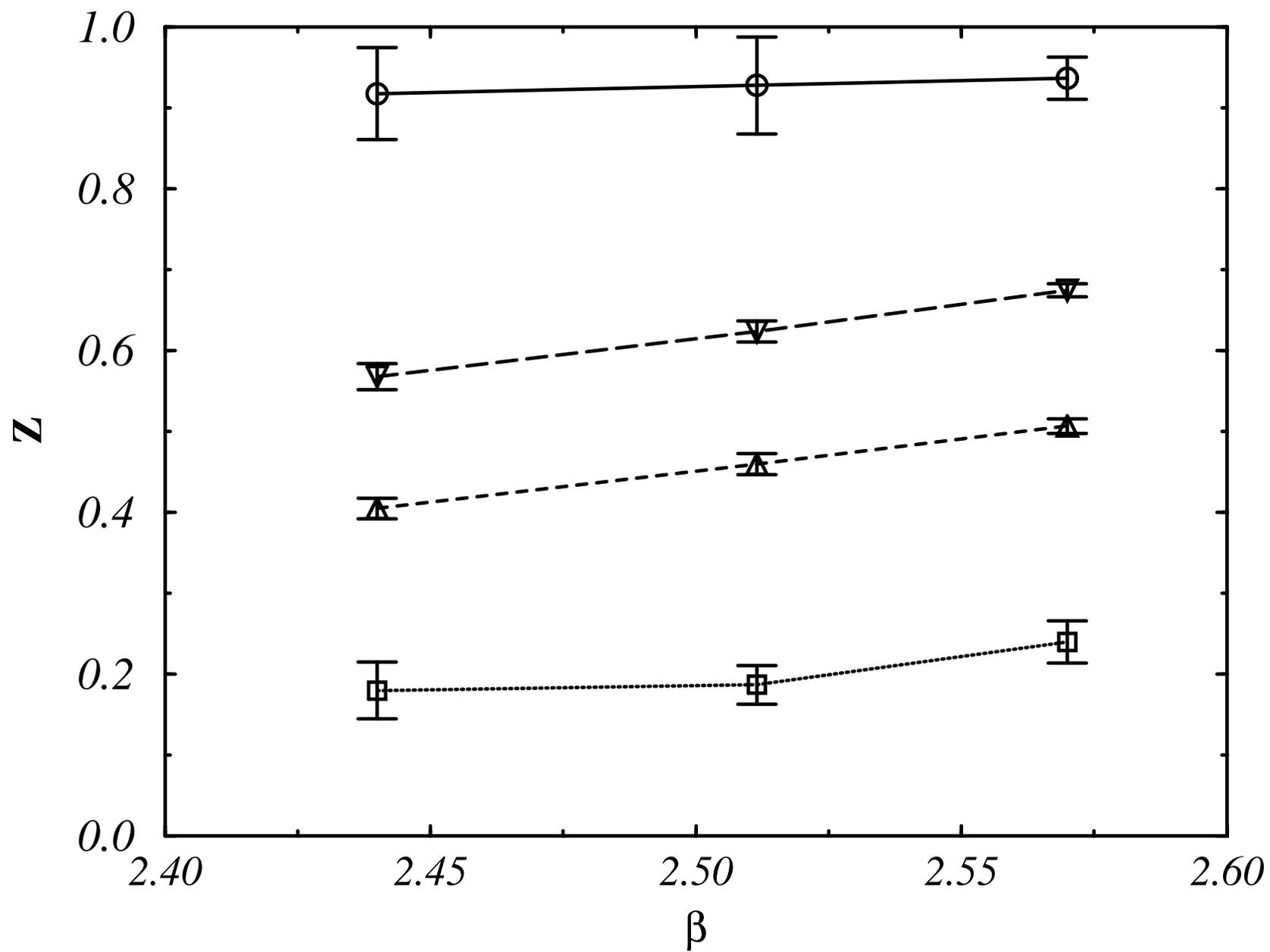
\item[Figure 1.] Values of $Z$ as obtained by the heating method
for the geometric and 0--, 1-- and 2--smeared field--theoretic topological 
charges, (circles, squares, up--triangles and down--triangles respectively).

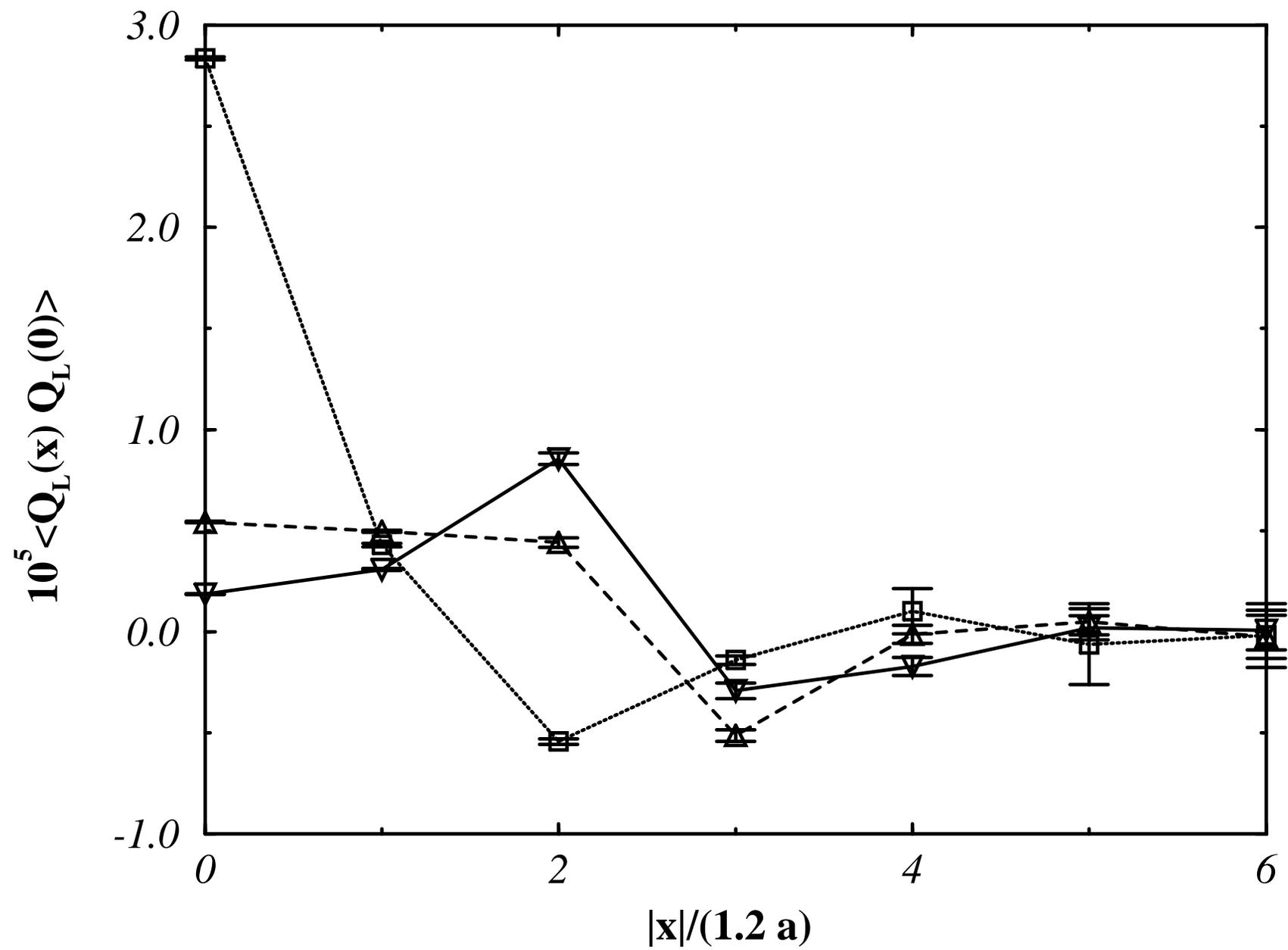
\item[Figure 2.] Correlation function $\langle Q_L(x) Q_L(0) \rangle$
as a function of $|x|/(1.2\;a)$ for the 0--, 1-- and 2--smeared topological 
charges (squares, up--triangles and down--triangles respectively) 
at $\beta=2.57$.

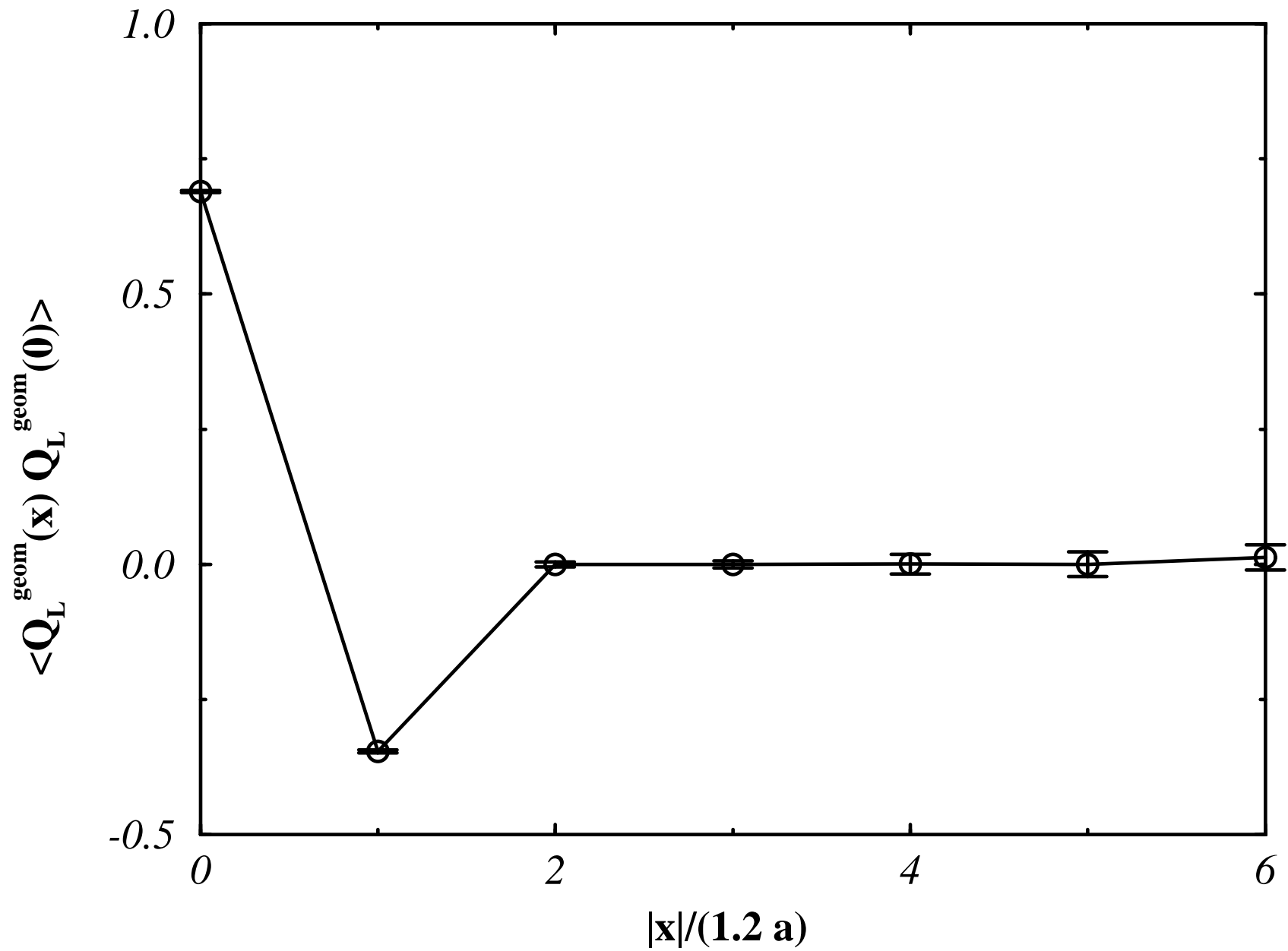
\item[Figure 3.] The same as in Fig.~2 for the geometrical topological
charge.

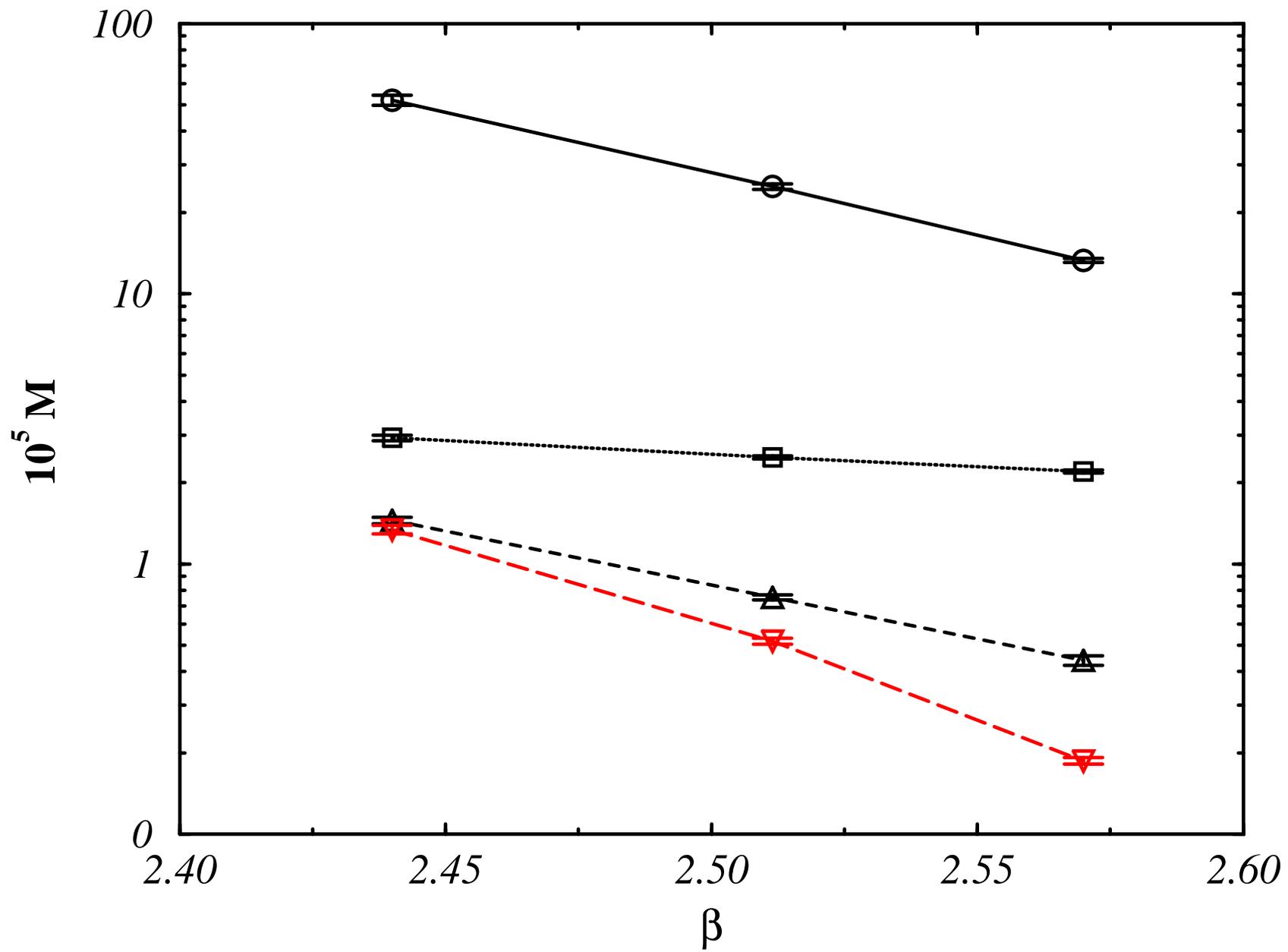
\item[Figure 4.] Values of $M$ as obtained by the heating method
for the geometric and 0--, 1-- and 2--smeared field--theoretic topological 
charges, (circles, squares, up--triangles and down--triangles respectively).

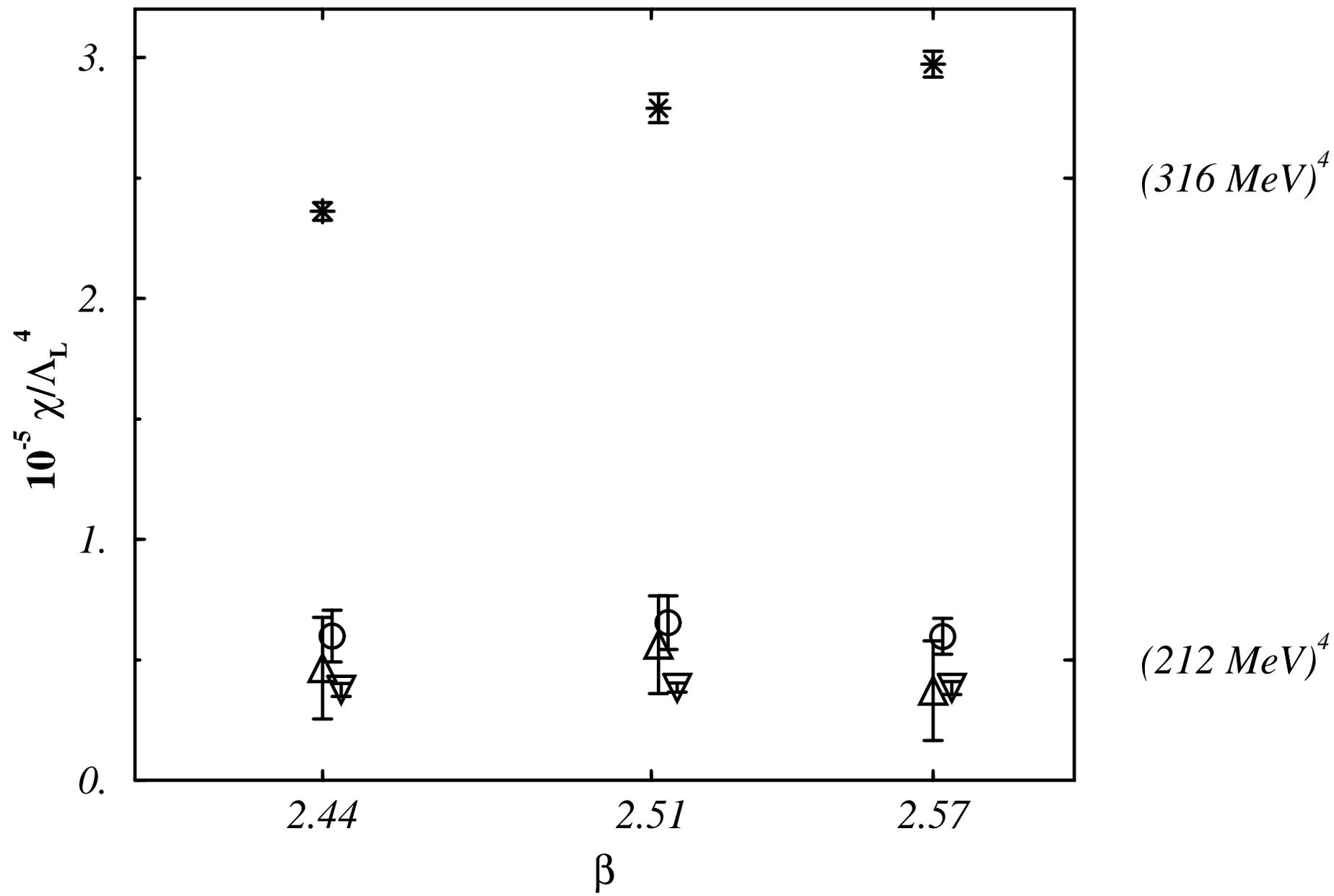
\item[Figure 5.] $\chi$ in $\Lambda_L$ and MeV units for the 
unsubtracted geometrical charge (stars), subtracted geometrical
charge (circles) and 0-- and 2--smeared charge (up and down triangles).

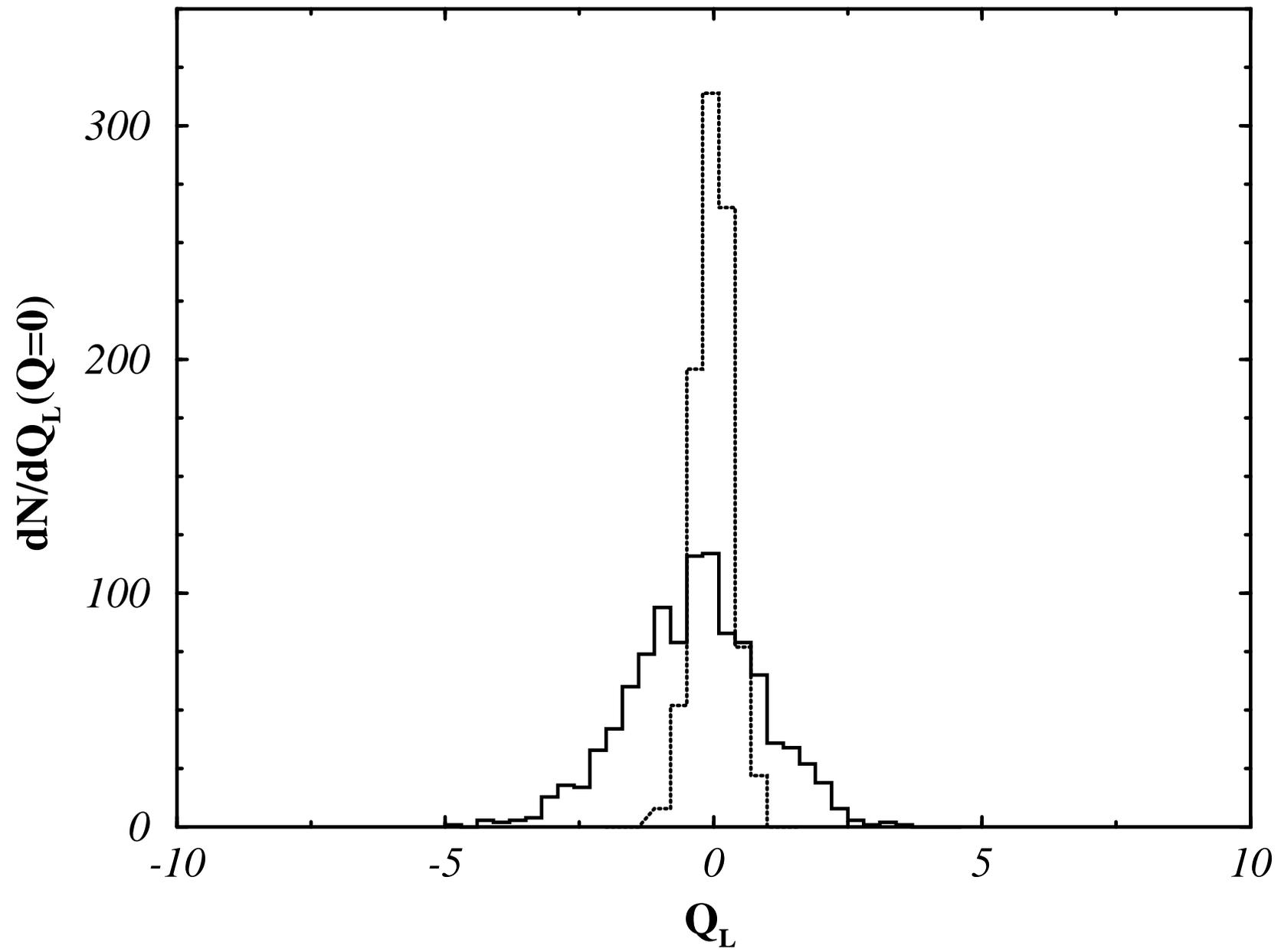
\item[Figure 6.] Distribution of $Q_L$ in the zero--topological charge sector
$Q=0$ for the 0--smeared (solid line) and 2--smeared (dotted line) topological
charge densities at $\beta=2.57$. 

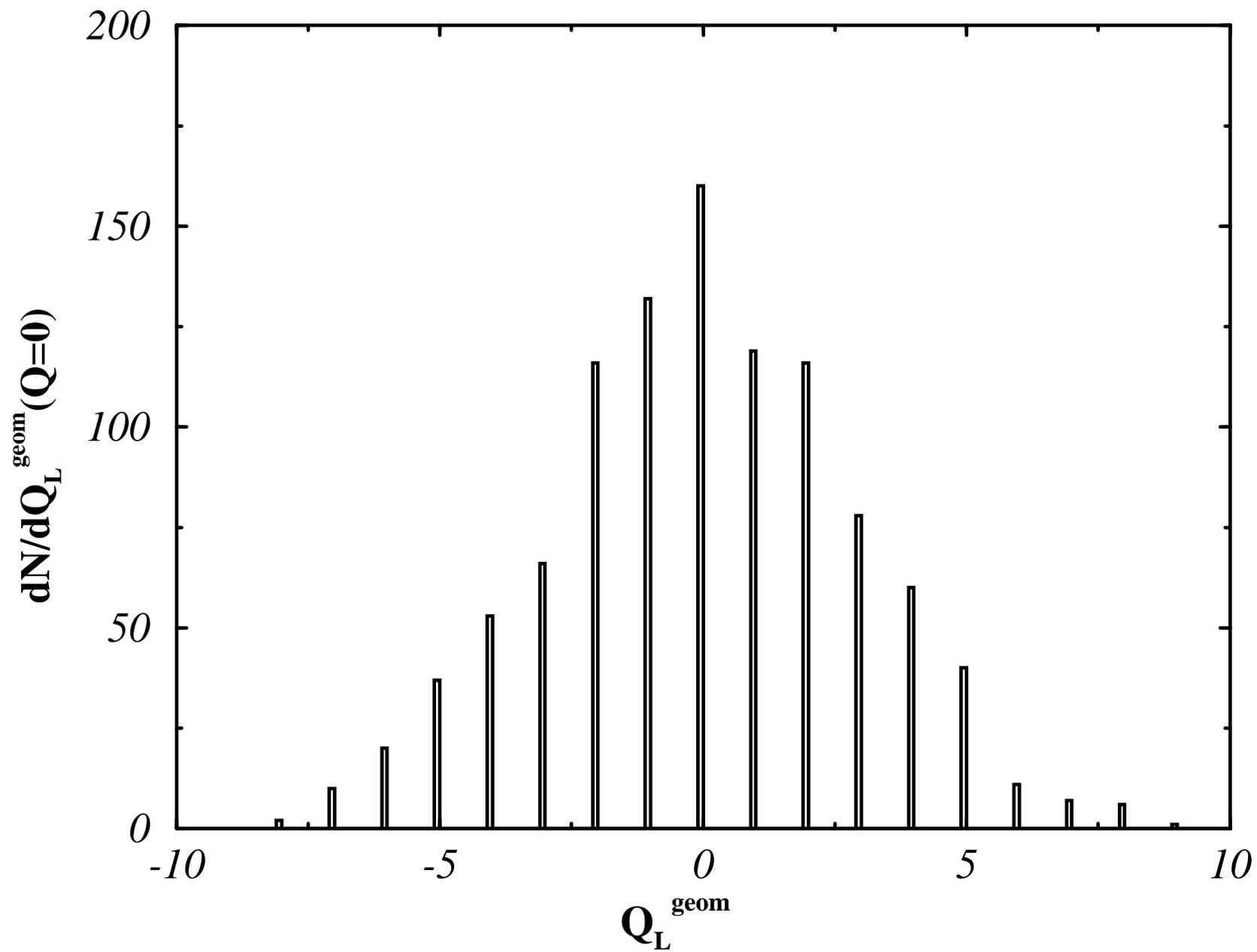
\item[Figure 7.] The same as in Fig.~6 for the geometrical topological
charge.

\end{enumerate}

\vskip 2cm

\noindent{\bf Table caption}

\vskip 5mm

\begin{enumerate}

\item[Table I.] $\chi_L$, $Z$ and $M$ for the 0--, 1--, 2--smeared and
geometric topological charge density operators at $\beta=2.57$.

\end{enumerate}

\newpage

\vskip 2cm

\centerline{\bf Table I}

\vskip 5mm

\moveright 0.1 in
\vbox{\offinterlineskip
\halign{\strut
\vrule \hfil\quad $#$ \hfil \quad & 
\vrule \hfil\quad $#$ \hfil \quad & 
\vrule \hfil\quad $#$ \hfil \quad & 
\vrule \hfil\quad $#$ \hfil \quad \vrule \cr
\noalign{\hrule}
\hbox{operator} & 
10^{5} \times \chi_L &
Z &
10^5 \times M \cr
\noalign{\hrule}
\hbox{\sl geometric} & 16.6(3) & 0.937(26) & 13.26(23) \cr
\noalign{\hrule}
\hbox{\sl 0--smeared} & 2.320(52) & 0.240(26) & 2.200(32) \cr
\noalign{\hrule}
\hbox{\sl 1--smeared} & 1.010(49) & 0.507(9) & 0.440(18) \cr
\noalign{\hrule}
\hbox{\sl 2--smeared} & 1.165(64) & 0.675(8) & 0.187(5) \cr
\noalign{\hrule}
}}

\end{document}